\documentclass[aps,prl,showpacs, twocolumn]{revtex4} 
\usepackage{graphicx}
\usepackage{bm}
\usepackage{amsmath}
\usepackage{hyperref}

\providecommand{\ket}[1]{\lvert #1 \rangle}

\providecommand{\be}{\begin{equation}}
\providecommand{\ee}{\end{equation}}
\providecommand{\ba}{\begin{eqnarray}}
\providecommand{\ea}{\end{eqnarray}}

\usepackage{mathbbol}
\begin{document}

\title{Modeling Leggett-Garg-inequality violation}
 
\author{ S. V. Moreira $^1$, A. Keller$^2$, T. Coudreau$^1$ and P. Milman$^1$}

\affiliation{$^{1}$Univ. Paris Diderot, Sorbonne Paris Cit\'e, Laboratoire Mat\'eriaux et Ph\'enom\`enes Quantiques, UMR 7162, CNRS, F-75205, Paris, France}
\affiliation{$^{2}$Univ. Paris Sud, Institut des Sciences Mol\'eculaires d'Orsay (UMR 8214 CNRS), F-91405 Orsay, France}

\begin{abstract}
The Leggett-Garg inequality is a widely used test of  the ``quantumness" of a system, and involves correlations between measurements realized  at different times. According to its  widespread interpretation, a violation of the Legget-Garg inequality disproofs macroscopic realism and non-invasiveness. Nevertheless, recent results point out that macroscopic realism is a model dependent notion and that one should always be able to attribute to invasiveness a violation of a Legget-Garg inequality. This opens some natural questions: how to provide such an attribution in a systematic way? How can apparent macroscopic realism violation be recast into a dimensional independent invasiveness model? The present work answers these questions by introducing an operational model where the effects of invasiveness are controllable  through a parameter associated with what is called the {\it measurability} of the physical system. Such a parameter leads to different generalized measurements that can be associated with the dimensionality of a system, to measurement errors or to back action.
\end{abstract}
\vskip2pc 
 
\maketitle

\section{I. INTRODUCTION}
Most of the introductory sentences of scientific papers in the domain of quantum mechanics express its contrasts and contradictions with our everyday life-built intuition. One may wonder if what is at the heart of the quantum mechanical incoherences with our world is the elusive definition of what is ``classical". It is of course tempting to define as classical everything that does not seem to behave quantum mechanically, as for instance objects in human scale. Nevertheless, even using this intuitive notion poses problem, since it is not more straightforward to define what one means by ``behave quantum mechanically" than the opposite. One aspect of quantum mechanics that is incompatible with a classical theory, defined as local and realist by Einstein, Podolsky and Rosen \cite{EPR, Bohr}, is Bell-type inequalities violation \cite{Bell, CHSH, Aspect}. However, one cannot safely assert that all states that do not violate these inequalities are nonquantum. Perhaps the most general property of quantum mechanics is contextuality \cite{Cabello, Cabello2, KochenSpecker}, that can be observed for {\it any} quantum state \cite{Cabello3, Badzia}. Assuming noncontextuality allows one to derive inequalities, of which Bell's are a special case. Such inequalities suffer, in their broader and state independent version,  of a lack of intuitive interpretation. They do not classify quantum states according to any usefulness they may have as a resource, or any of their particular properties that help one's understanding of the quantum-classical frontier. In this spirit, Leggett and Garg \cite{LG} proposed, in the 1980s, an inequality that is often presented as enabling to witness quantumness in a macroscopic system when violated. This inequality, known as the Leggett-Garg inequality (LGI), involves measurements of observables that can be associated with macroscopic properties of a system at different times, defined here as $\hat Q(t_i)$. By defining  $C_{kl} \equiv \langle \hat{Q}(t_k)\hat{Q}(t_l)\rangle$, the LGI can be written as  \cite{Emary}:
\begin{equation}\label{eq:eq1}
-2 \le K_{LG} \equiv C_{12}+C_{23}+C_{34}-C_{14} \le 2.
\end{equation}
In the usual interpretation of the LGI, it must be satisfied if the following two assumptions hold: \\ 
{\it (i) Macroscopic Realism (MR):} a macroscopic system with two or more macroscopically distinct states available to it will at all times be in one of those states.\\
{\it (ii) Noninvasive measurability (NIM):} it is possible, in principle, to determine the state of the system with arbitrarily small perturbation to its subsequent dynamics.\\

Violation of the LGI would thus imply nonclassicality as defined above \cite{Budroni}. Experimental tests of the LGI and propositions can be found in \cite{Xu, Goggin, Athalye, Souza, Knee, Dressel, Emary2, Palacios-Laloy, Groen}, where \cite{Palacios-Laloy} is probably the first experimental violation of the LGI involving measurements realized in a quantum number associated with a macroscopic system. Nevertheless, the notion of ``macroscopicity"  in the LGI has always been somewhat controversial. Since one of the mathematical requirements in its derivation is that correlations between measurements should be correlations between dichotomic observables, one may wonder what is the sense of macroscopicity.  Can one talk of {\it macroscopicity} when only two effective quantum numbers are assigned to each observable? In \cite{Koffler}, some physical insight was provided to help interpret the role of macroscopicity in LGI: it was shown that classical physics emerges when the  resolution of a classical measurement apparatus decreases faster with the dimensionality of a system than the intrinsic quantum noise increases. In this case, the dimensionality of the system is associated with the value of the total spin $j$. Macroscopicity emerges in the limit of  $j \rightarrow \infty$, for which measurements become coarse grained and, consequently,  noninvasive. 

\section{II. WHAT DOES THE LGI TEST?}
Recently, another issue was raised,  expressing some discontent of the community with the usual interpretations of a LGI violation \cite{Maroney, Clemente}. It was shown in Ref. \cite{Maroney}, that the definition of macroscopic realism appearing in (1) is model dependent, since in order to be tested, it assumes the  Copenhagen interpretation of quantum mechanics. In the authors' point of view, which is the one assumed here, macroscopic realism, as usually interpreted in LGI's is analogous to a  superselection rule \cite{Zurek} that prepares the system in a statistical mixture of some privileged states. According to this definition, violating LGI would imply that the system is, at some point, in a quantum superposition. However, the notion of quantum superposition is not necessarily incompatible neither with macroscopicity nor with realism \cite{Bohm}.  The authors conclude thus that the only assumption that is actually being tested in a LGI in a model independent way is invasiveness.

This approach of the LGI becomes even clearer if one considers its mathematical formulation, which makes explicit what features of correlations between measurement results are being tested. To this end, one should assume that observables $\hat Q (t_i)$ are measured, at each one of the many runs of an experiment, in order to compute the statistical average needed to obtain the correlations $C_{kl}$. When the two-time correlations $C_{kl}$ are computed in the situation where the four measurements $\hat Q (t_i)$, $(i=1,2,3,4)$, were realized,  inequality \eqref{eq:eq1} is always satisfied, both in the classical and in the quantum realm. The important point is when one considers the case where only the two measurements $\hat Q (t_k)$ and $\hat Q (t_l)$ are made at each run to compute $C_{kl}$. In this case, Eq. (\ref{eq:eq1}) is valid only if the noninvasiveness assumption is made. This last situation (two measurements at each run) corresponds to the LGI. Thus, the LGI tests the pertinence of the hypotheses that correlations between two measurement outcomes realized at different times are undisturbed by the realization of  other measurements to the system at different times. Given these results, how can one interpret previously obtained ones, where macroscopicity seemed to enhance coarse graining, leading to a quantum to classical transition and the loss of violation of the LGI? Is it possible to model and control invasiveness in a way that does not depend on the system's size? 

\section{III. MODELING INVASIVENESS OF MEASUREMENTS IN LGIs}
 With the purpose of formalizing the previous discussion and determining in a precise mathematical way the interplay between invasiveness, dimensionality and violation of the LGI, we propose, in the present manuscript,  a mode to test a LGI using positive operator valued measurements (POVM).  In our model, invasiveness can be associated with the resolution and efficiency of a measurement apparatus.  These parameters, according to the experimental situation, may, or not, be associated with the dimensionality of the system. The introduced POVMs  provide a physically sound and operational interpretation of what is actually being tested by the LGI, unifying and clarifying the notions of invasiveness and macroscopicity. 

The system considered consists of a spin $j$,  associated, for instance, with an atom or, equivalently, an orbital angular momentum.  Therefore, $\hat{J}_{\alpha}$ , $\alpha=x, y, z$, is the $\alpha$-component of $\hat{{\bf J}}$, the total spin operator. $\hat J_z$ eigenstates are denoted as $\ket{m}$, $-j \leq m \leq j$. We consider that the dynamics of the system is governed by the Hamiltonian:  
\begin{equation}\label{eq:eq3}
\hat{H}=\Omega \hat{\bf{J}}^2+\omega\hat{J}_x,
\end{equation}
where $\Omega$ and $\omega$ are constants  with the dimension of frequency.

We then define measurements that help clarifying the conditions for violating the LGI (\ref{eq:eq1}) by manipulation of a controlled parameter. Such measurements are performed on an initially maximally mixed state
\begin{equation}\label{eq:eq2}
\rho(0)\equiv \frac{1}{2j+1}\sum_{m=-j}^j\left|m\right>\left<m\right|
\end{equation}
to ensure that nonclassicality can only appear from the system's dynamics and the subsequent measurements.  In \cite{Koffler}, it was shown that defining $\hat Q (0)=\hat \Pi_z$, where $\hat \Pi_z=\sum_m(-1)^{j-m} \left|m\right>\left<m\right|$ is the parity operator, leads to violation of a LGI irrespectively of the dimensionality of the system (value of $j$) and for an initial state as (\ref{eq:eq2}). Parity is a projective measurement that involves all the states $\ket{m}$, creating the analogous of ``collective states" associated with a single quantum number. Parity reduces systems of any dimension to an effective two level system through a mapping that applies to all possible states. Thus, the time evolution of the parity operator involves all the system's states in the same way (except for a sign change). It is thus intuitively acceptable that such a measurement will always lead to a LGI violation, since one can hardly think of a less invasive dichotomic measurement. 

Our strategy consists  in defining an observable depending on a parameter $\sigma$ that, at its extremal value, leads to the parity operator. In order to do so, we recall that an important ingredient in the LGI inequality is that correlations $C_{kl}$ are as binary ones, {\it i.e.}: 
\begin{equation}\label{eq:eq6}
C_{kl}\equiv p_+^{kl}q_{+|+}^{kl}+p_-^{kl}q_{-|-}^{kl}-p_+^{kl}q_{-|+}^{kl}-p_-^{kl}q_{+|-}^{kl}.
\end{equation}
where $p_\pm^{kl}$ are the probabilities of measuring one of the outcomes $\pm 1$, and $q_{\pm|\pm}^{kl}$ are the probabilities of obtaining the $\pm$ outcomes conditioned to what was previously obtained. Using this fact, Asadian {\it et al.} proposed LGI test using the measurement of periodic observables defined in an nanomechanical oscillator \cite{Asadian} and in \cite{Ketterer} we showed that Bell type inequalities can be performed using observables with an arbitrary spectrum. In both results, the key ingredient is defining a two-valued POVM $\hat M_{\pm}$, as follows:

\begin{equation}\label{eq:2}
\hat{E}_{\pm}=\hat{M}_{\pm}^\dagger\hat{M}_{\pm}=\frac{1}{2}(1\pm\hat{A}),
\end{equation}
where $\hat A$ is an operator with a spectrum in the interval $[-1,1]$.
Using Neumark's theorem\cite{Peres}, we have that each element of this POVM, identified by the signs $\pm$,  is  associated with one of the two possible outcomes $(\pm 1)$ of a projective measurement realized in an auxiliary two dimensional Hilbert space. Operators $\hat{A}$  have thus a spectrum bounded between $\pm 1$, since $\langle \hat{A}\rangle=P_+-P_-$ and $P_++P_-=1$,  where $P_\pm$ are the probabilities to obtain one of the two possible  outcomes of measurements realized in the auxiliary two dimensional space. 

\begin{figure}[h]

\centering 
\includegraphics[width=8.5cm]{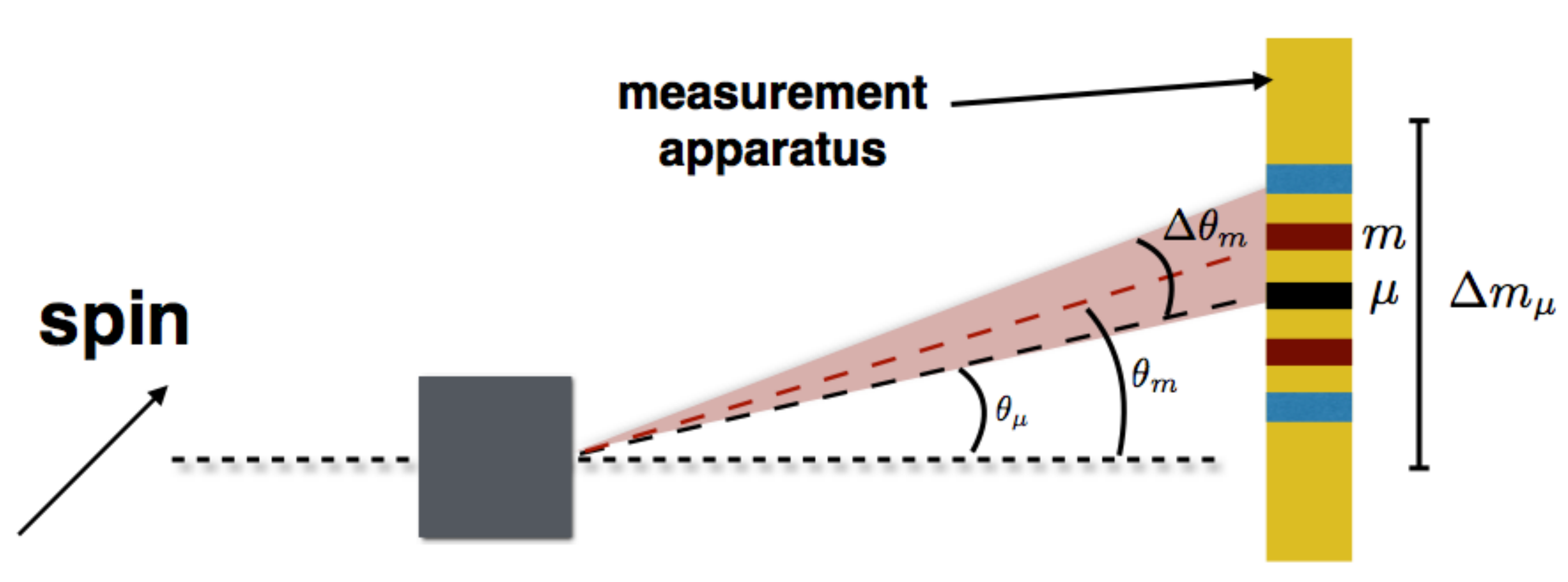} 
\caption{Illustration, in the context of a Stern Gerlach type experiment, of the introduced parameters governing the two valued POVM in Eq. (\ref{eq:eq10}): a spin $j$ is deflected by a magnetic field with spatial inhomogeneities. As a consequence, for some values of $m$ ($m=\mu$), the perfect measurement of $m$ is possible (perfect correlation between $m$ and the deflection angle, $\theta_{\mu}$, and the position on the screen, the measurement apparatus). For $m \neq \mu$, uncertainties $\Delta \theta_m$  around  the deflection angle $\theta_m$ lead to uncertainties in the position of the spin in the measurement apparatus, and consequently, in the value of $m$.  The interval $\Delta m_{\mu}$ is the resolution of the measurement, in the sense that it determines the number of $\mu$'s that can be faithfully measured ({\it i.e.}, are associated to a fixed position in the screen). }\label{fig1}

\end{figure}

\begin{figure}[h]

\centering 
\includegraphics[width=8.2cm]{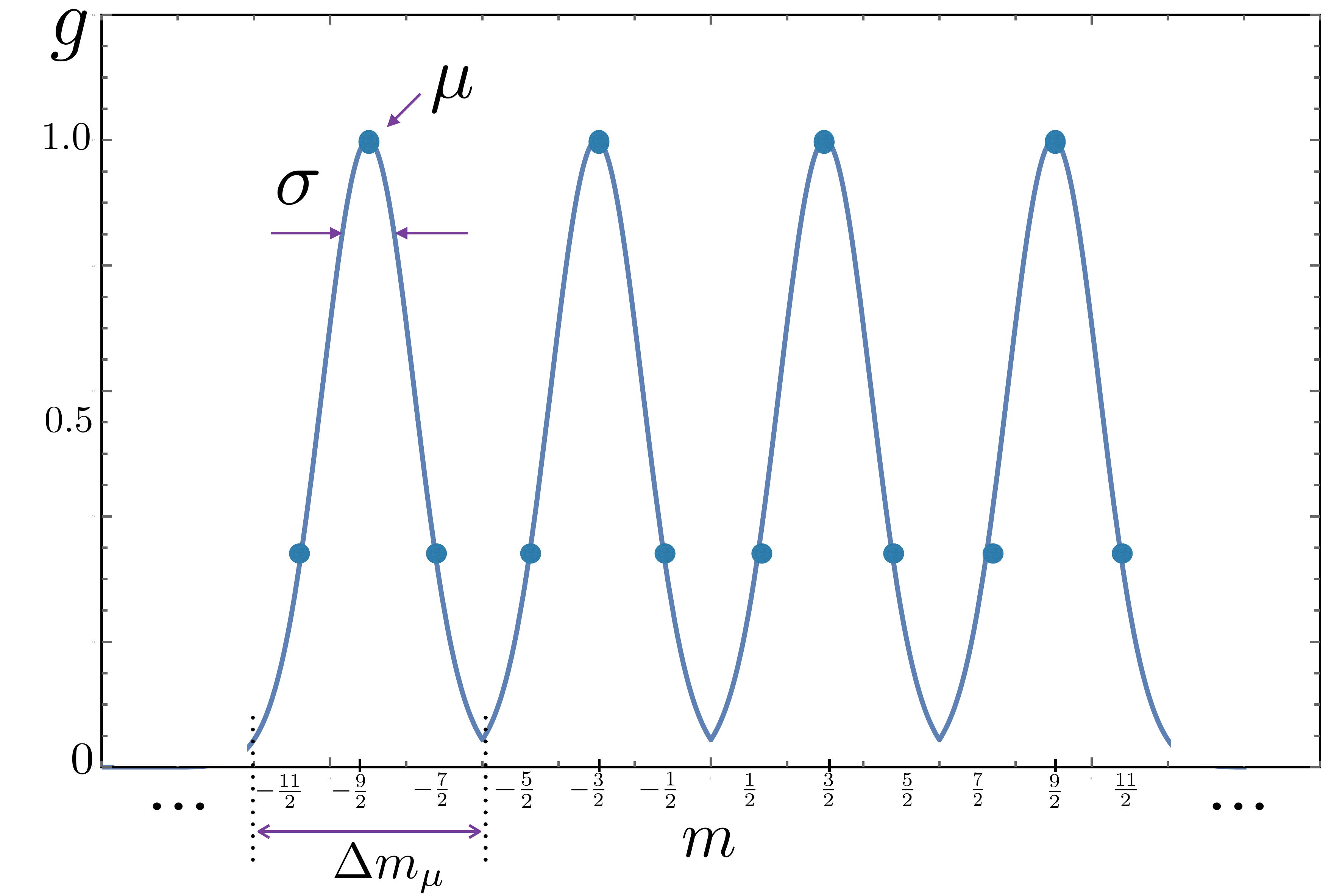} 
\caption{ Representation of the function $g$ and definition of the parameters of the POVM in Eq. (\ref{eq:eq10}) as a function of $m$: the measurement results $\mu$ are the central values of $m$ in the interval $\Delta m_{\mu}$. For $m=\mu$, $g=1$. The width $\sigma$ determining the measurability is associated to the width of the gaussian function defined in each interval $\Delta m_{\mu}$. Nevertheless, we can notice that it is not the usual variance, since irrespectively of its value, $g=1$ for $m=\mu$. In the Figure, $\sigma=0.6$ and the number of elements in $\Delta m_{\mu}$ is given by ${\cal N}( \Delta m_{\mu})=3$. }\label{fig:fig2}
\end{figure}

We now introduce the following operator, specifying (\ref{eq:2}), that gathers all the required previously mentioned qualities and properties, and whose parameters are illustrated in Figs. \ref{fig1} and \ref{fig:fig2} and explained below: 
\begin{equation}\label{eq:eq10}
\hat{A}\equiv  \sum_{\mu}\sum_{m \in \Delta m_{\mu}}(-1)^{(j-m)}f_{{\mu}}(m, \sigma)\left|m\right>\left<m\right|,
\end{equation}
where $f_{{\mu}}(m, \sigma)=e^{\frac{-(m-\mu)^2}{2\sigma^2}}$ and $\Delta m_\mu$ are disjoint sets containing equally sized intervals of $m$. Operator (\ref{eq:eq10}) can be seen as an imperfect parity measurement. If  $f_{{\mu}}(m, \sigma)=1 \ \forall m$, $\hat A$ becomes the parity operator, and $\hat E_{\pm}$ are projective measurements. In this case, to each $m$ one assigns a $\pm 1$ eigenvalue, according to the parity of $j-m$. We now see how to interpret $\hat A$ in the general case. This will be done by providing a physical meaning to the parameters $\sigma$, $\mu$ and $\Delta m_{\mu}$. They will be associated with the concepts of  {\it measurability}, {\it optimal measurement}, and {\it measurement resolution}, respectively. 

We start by interpreting $\Delta m_{\mu}$ in Eq. (\ref{eq:eq10}), the measurement resolution. It determines the number $N$, among all the possible values of $m$, that the measurement apparatus can faithfully detect. By this, one means: detect the particle and assign to it the correct value of $m$. Each of these values is defined as $\mu$, the optimal measurement. We have a total of $N=(2j+1)/{\cal N}(\Delta m_{\mu})$ different values of $\mu$ for a spin $j$ system, where ${\cal N}(\Delta m_{\mu})$ is the number of elements in $\Delta m_{\mu}$. We see, in Fig. \ref{fig1}, that this corresponds to a perfect Stern-Gerlach type measurement: the $z$ axis projection of  the spin with value $m=\mu$ is deflected by an angle $\theta_m=\theta_{\mu}$ that allows one to identify its hitting position to $m$ univocally. In Eq. (\ref{eq:eq10}), it translates as: for $m=\mu$, $f_{{\mu}}(m, \sigma)=1$ irrespectively of the value of $\sigma$. Thus, a correct and well defined parity is associated with this values of $m$, {\it i.e.}, $\langle \hat A \rangle_{m=\mu}=(-1)^{(j-m)}=\langle \hat \Pi_z \rangle_{m=\mu}=\pm 1$.  The measurement resolution $\Delta m_{\mu}$ also defines the interval of values of $m$ that are considered around each $\mu$: each interval $\Delta m_{\mu}$ defines the domain of a function $f_{{\mu}}(m, \sigma)$.  

We now define the parameter $\sigma$, the {\it measurability}. Later, we will see that it can be related to the invasiveness. $\sigma$ is a measure of the unfaithfulness of the measurement. By unfaithfulness, one means the following scenario: for finite $\sigma$ and $m \neq \mu$, the particle is detected, but the value of $m$ cannot be perfectly determined. Thus, it will be sometimes associated with the correct value, or to a value with the same parity (and then the correct parity will be assigned to it) and sometimes associated with a $m'$ with different parity. As a consequence, for $m\neq \mu$, one cannot assign a well defined parity, and $\langle \hat A \rangle_{m \neq \mu}= (-1)^{(j-m)}e^{\frac{-(m-\mu)^2}{2\sigma^2}}$. Notice that, in particular, for $\sigma$ such that $f_{\mu}(\mu+1, \sigma)=0$, $\langle \hat A \rangle_{m \neq \mu}=0$ and $\langle \hat E_{\pm} \rangle = 1/2$: we have the equivalent of a perfectly random parity measurement. Physically, unfaithfulness in the measurement could arise, in the context of a Stern-Gerlach type experiment, as a consequence of position dependent fluctuations of the magnetic field, that can create an uncertainty $\Delta \theta_m$ in the deflection angle depending on the value of $m$. All these parameters are illustrated in Fig. \ref{fig:fig2}, where we plotted $g \equiv f_{\mu}(m, \sigma)$, with ${\cal N} (\Delta m_{\mu})=3$, $\sigma=0.6$ and one has three different values of  $\mu$ for the considered  system. 

We conclude this discussion by noticing that the limit of $\sigma \rightarrow \infty$ (perfect measurability) is equivalent to the one of perfect resolution ($f_{\mu}(m,\sigma)=1 \forall m$, ${\cal N}(\Delta m_{\mu})=1$), since all the values of $m$ are associated to a $\mu$, and  $N=2j+1$. 

We now move to an example that  illustrates how the introduced parameters can control the invasiveness of a measurement in a way that is dimension and model independent. We then discuss how they can be modified, depending on the experimental situation, by the system's dimensionality or measurement efficiency. To this effect, we notice that the probabilities appearing in Eq.   (\ref{eq:2}) can be written, for a given pair of measurement times $t_k, t_l$, as:
\begin{eqnarray}
&&p_+^{kl}= Tr[E_+\hat{\rho}(t_l)], \nonumber \\
&&q_{-|-}^{kl}=Tr[E_+\hat{\rho}_\pm(t_k)],
\end{eqnarray}
where we have used Eq. (\ref{eq:eq3}),  with $\hat{U}( t_k-t_l)=e^{-i\theta_{kl}\hat{J}_x}$, with $\theta_{kl}=\omega (t_k-t_l)$. Thus,
\begin{equation}
\hat{\rho}_\pm(t_l)=\hat{U}(t_k-t_l)\hat{\rho}_\pm(t_k)\hat{U}^\dagger(t_k-t_l t)=e^{-i\theta_{kl}\hat{J}_x}\hat{\rho}_\pm(t_k) e^{i\theta_{kl}\hat{J}_x},
\end{equation}
with
\begin{equation}
\hat{\rho}_\pm(t_k)=\frac{\hat{M}_\pm\hat{\rho}(t_k)\hat{M}^\dagger_\pm}{p_{\pm}}=\frac{1\pm \hat{A}}{2(2j+1)p_{\pm}}.
\end{equation}
Using  \cite{Curtright}, correlations $C_{kl}$ can be computed, and we will do so in a simple illustrative example of a spin $j=5/2$ (a six level system). In order to be able to identify the above introduced parameters, we will consider that there are only two possible values of $\mu$, $\mu_{\pm}=\pm 5/2$, and $\Delta m_{\mu_{\pm}}=[\pm 5/2, 0]$ (intersection between the two sets is unimportant, since $0$ is not a possible value of $m$).

In this simple example, we can identify three possible values of $f_{\mu_{\pm}}(m,\sigma)$, $a$, $b$ and $c$, that obey the following relations: $a=1$ and $c=b^4$, with  $b=e^{-1/2\sigma^2}$, since the $f_{\mu_{\pm}}(m,\sigma)$ considered here is a gaussian distribution centered at $\mu_{\pm}$. 
 
\begin{figure}[h]

\centering 
\includegraphics[width=8.2cm]{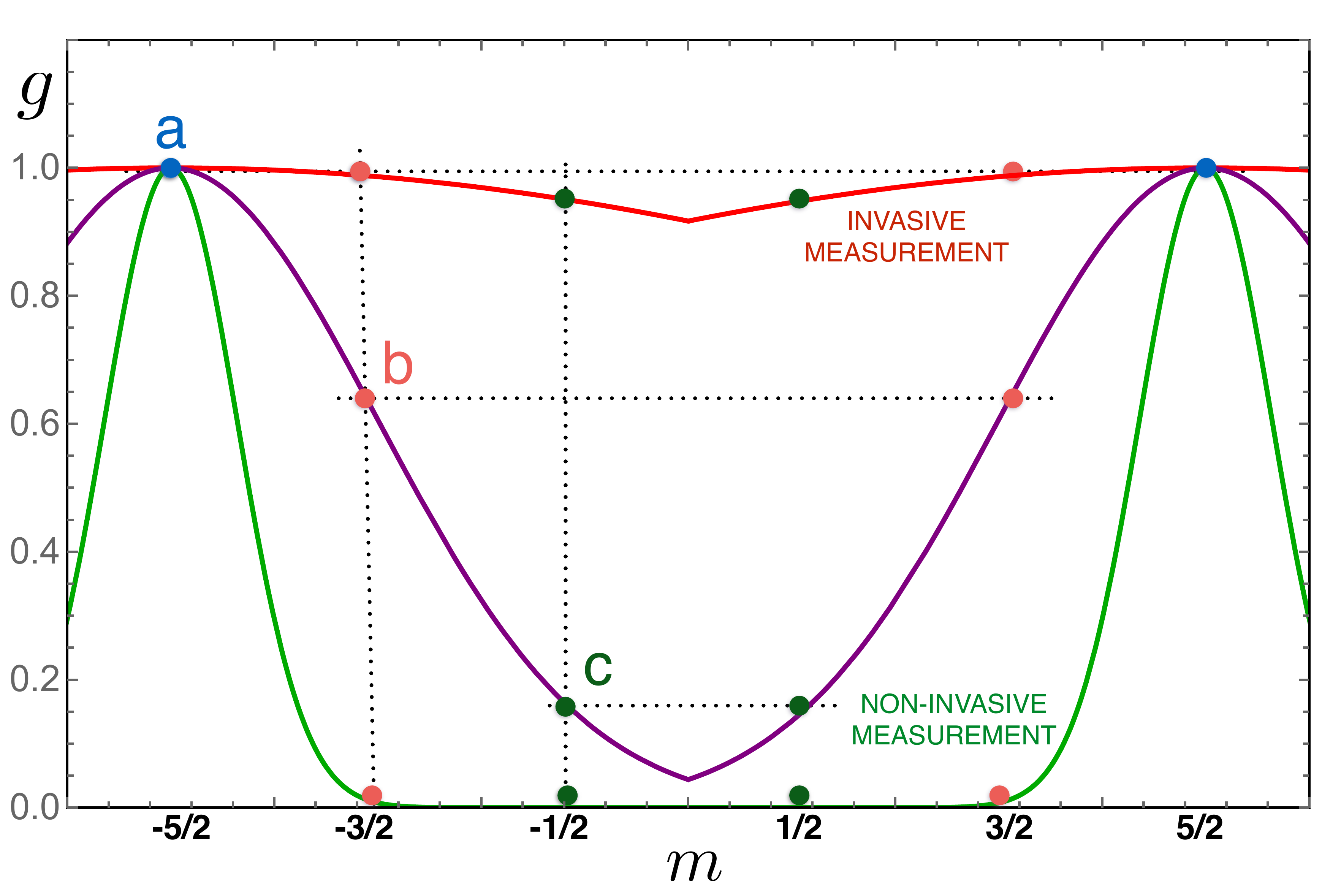} 
\caption{ Representation of the three possible values of $g$ ($a$, $b$ and $c$) in the illustrative case of  $j=5/2$ as a function of $\sigma$ for the case where  $m_{\pm}={\pm} 5/2$ are optimally measured. The parameters $a$, $b$ and $c$ are represented for three possible choices of $b$ (and consequently $c$): red, $b=0.98$ is associated to a invasiveness measurement, and high measurability; purple, $b=0.61$ and green, $b=0.008$  are associated to non invasiveness measurements (from up to bottom curve, respectively). } \label{fig3}
\end{figure}

It is clear that modifying $\sigma$ leads to a modification of $b$ and $c$ only ($a$ is constant, as previously defined). Thus, for $\sigma \ll 1$, only $\mu_{\pm}=\pm 5/2$ are faithfully measured while for $\sigma \rightarrow \infty$ the operator $\hat A$ tends to a parity measurement.

\begin{figure}[h]

\centering 
\includegraphics[width=9.2cm]{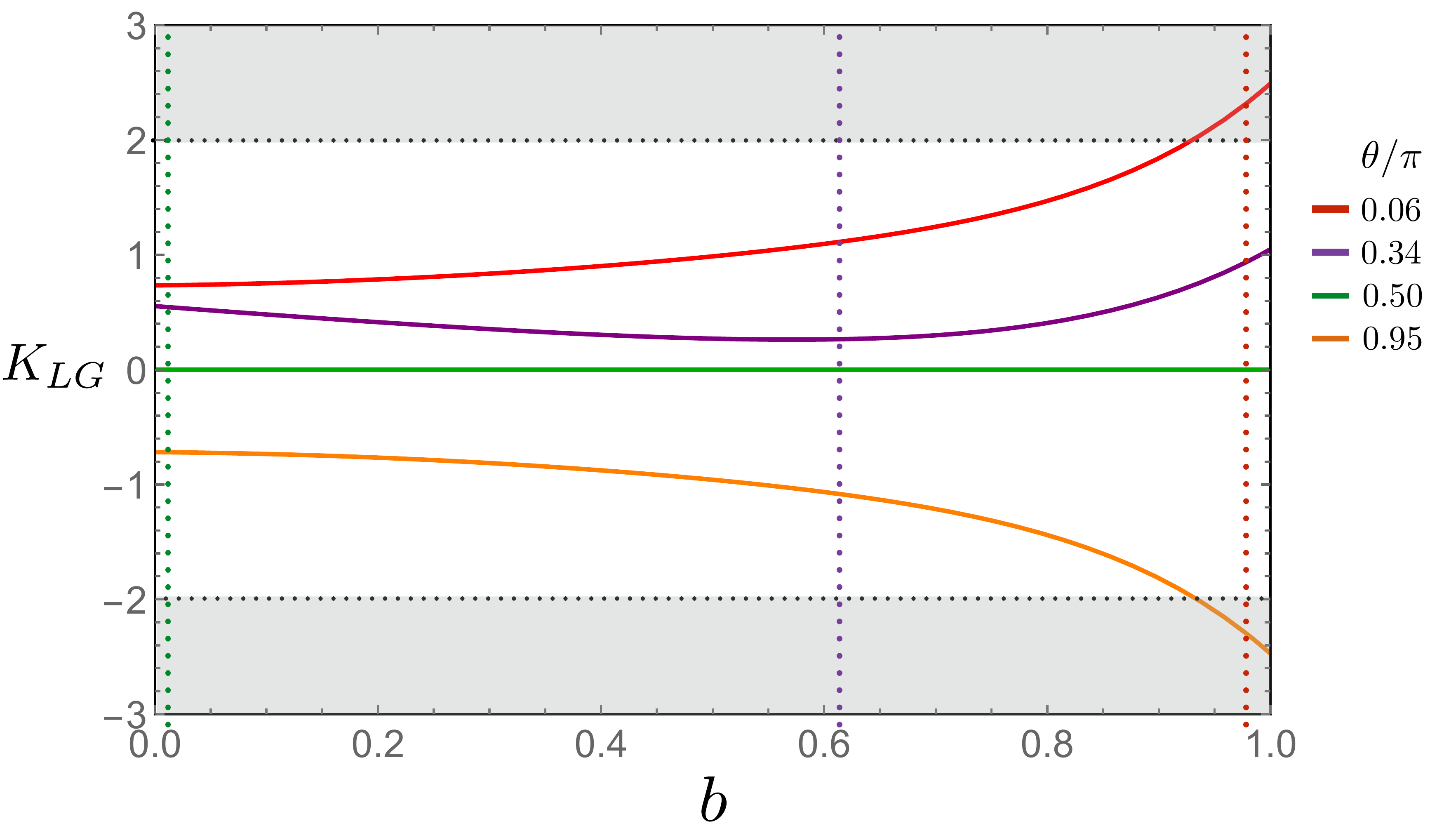}
\label{fig:fig14}
\caption{Legget-Garg parameter, $K_{LG}$ (Eq. \eqref{eq:eq1}) as a function of $b$ for the following values of $\theta/\pi$ (from up to bottom curve, respectively): $0.06$ (red), $0.34$ (purple), $0.50$ (green), $0.95$ (orange). We identified, for the cases where violation is observed for some value of $b$ (red and orange curves, corresponding to $0.06$ and $0.95$ as values for $\theta/\pi$, respectively) the values of $b$ corresponding to the functions $g$ plotted in Fig. \ref{fig3}. They are represented by dotted vertical lines: red for  $b=0.98$, purple for $b=0.61$ and green for $b=0.008$. The chosen color code is the same as in Fig. \ref{fig3}.\label{fig:cut}}
 
\end{figure}

We computed $K_{LG}$ as a function of $b$ (which here, equivalently to $\sigma$, is related to the measurability) and of time. We considered, for simplifying and illustrative reasons, only measurement times such that  $k=l+1$ and define $\theta_{l+1l} \equiv \theta$. Results are plotted in Fig. \ref{fig:cut}. One can notice that the absolute value of $K_{LG}$ increases monotonically with $b$ for the cases where the inequality is violated for some $b$. Nonetheless, its maximum is always reached for $b=1$, which corresponds to $\sigma \rightarrow \infty$. In this situation, we retrieve the parity operator $\hat \Pi_z$, and we have perfect measurability and measurement resolution.  

We can thus associate invasiveness to measurability: the more a system is measurable, in  the sense that the more one can faithfully detect different values of $m$ (higher value of $b$), the more the LGI is violated.  Measurability is a dimension independent definition, but it can perfectly well depend on the dimensionality of a system in a similar way as in  \cite{Koffler}. By making $\Delta m_{\mu}$ increase faster with $j$ than $\sigma$, one can, with increasing $j$, lose violation of the LG inequality. In the present measurement model, this can be understood easily as a increase of the measurability that is slower than the the increase in resolution (increase of $\sigma$ slower than of $\Delta m_{\mu}$). Nevertheless, such a behavior can be observed irrespectively of the dimension of the system. Finally, one should notice that our model can also be interpreted as a measure of disturbance of a measurement: unfaithful measurements can also be modeled by a measurement that highly disturbs the system, {\it i.e.}, modifies the value of $m$ after measuring it. Consequently,  a value of $m$  is assigned to a state but, after the measurement process, it is no longer applicable. In this context, our model easily connects the introduced parameters to the notion of classical back action and disturbance of a measurement. The more classical back action is introduced, the less the LGI is violated, excluding the possibility of the clumsiness loophole \cite{Wilde}. 

\section{IV. CONCLUSION}
As a conclusion we provided a toy model where the violation of the LGI can be directly controlled by and understood through  physically sound parameters. These parameters can be associated to the unfaithfulness of a measurement,  a notion that can have different physical origins, all of them contemplated in the introduced model.  One example is the  increase of dimensionality of the system and another one is  the classical disturbance created by the measurement process. While each parameters' precise interpretation depends on the physical system one uses to test a LGI, the role of each parameter is clearly identified and related to the invasiveness, whatever its physical origin is.  Thus, this notion is shown to be, operationally, the most fundamental one that is tested in a LGI. Our model can be used to help understanding and interpretation of  LGI tests, and can be tested experimentally in a number of physical systems, as Stern-Gerlach like experiments with inhomogeneous fields, or the orbital angular momentum of photons.  

\section{V. ACKNOWLEDGMENT}
S. V. M. acknowledges financial support from the Brazilian agency CAPES.

\end{document}